\documentclass[twocolumn,prd,showpacs,amsmath,amssymb]{revtex4}

\usepackage{bm}

\begin{document}

\title{Li{\' e}nard-Wiechert Potentials of a Non-Abelian Yang Mills Charge}

\author{{\" O}zg{\" u}r Sar{\i}o\u{g}lu}
\email{sarioglu@metu.edu.tr}
\affiliation{Department of Physics, Faculty of Arts and  Sciences,\\
             Middle East Technical University, 06531 Ankara, Turkey}

\date{\today}

\begin{abstract}
Given the path of a point particle, one can relate its acceleration
and, in general, its kinematics to the curvature scalars of its trajectory.
Using this, a general Ansatz is made for the Yang Mills connection
corresponding to a non-Abelian point source. The Yang Mills field
equations are then solved outside the position of the point source
under physically reasonable constraints such as finite total energy
flux and finite total color charge. The solutions contain the Trautman
solution; moreover two of them are exact whereas one of them is found
using a series expansion in $1/R$, where $R$ is the retarded distance.
These solutions are new and, in their most general form, are not gauge
equivalent to the original Trautman solution.
\end{abstract}


\maketitle

\section{\label{intro} Introduction}

In classical Maxwell electrodynamics, it is well known that accelerated charges
emit electromagnetic radiation. Using Li{\' e}nard-Wiechert (LW) potentials, one
is able to calculate the field strength for a point charge (or a system of point
charges) and relate the radiation from the accelerated charge to its motion and
the geometry of its trajectory \cite{bar}, \cite{jack}. One may, of course, ask
the same question in the case of non-Abelian Yang Mills (YM) theory, i.e. the
existence of a LW potential such that the emitted radiation and the trajectory
of the charge are interrelated. An analysis of this problem was done long ago
\cite{traut} and it was found that the color of a single point particle
remains constant although there is a transfer of energy, provided that the gauge
group is compact and semi-simple. Later it was also shown that the total gauge-invariant
color of an external line source in YM theory could change as a result of color
radiation, depending on the YM waves considered \cite{olt}.

Here we want to examine whether one can generalize the LW potential of \cite{traut}
in the light of the recent results we found relating the generalized acceleration
scalars $(a, a_{k})$ of a point particle to the curvature scalars of its trajectory
\cite{gs1}. To this effect, we make a new Ansatz for the YM vector potential
and find new solutions to the source-free YM field equations. Two of the solutions
we present are exact solutions whereas one of them is found using a series expansion
in $1/R$, the reciprocal of the retarded distance $R$. The solutions are obtained by
using physically acceptable constraints such as demanding that the total energy
flux $N_{YM}$ and the total color charge $Q$ are finite at large
$R$ values. We also examine whether the solutions presented are gauge equivalent
to the Trautman solution and show that they are not in their full generality.

In Section \ref{curve}, we give a brief review of some basic elements regarding
the geometry of a trajectory curve. A new Ansatz for the YM connection is
presented in Section \ref{enflux}. The total energy flux of the YM field
is defined and its behavior at large distances is examined. Demanding the
flux to be finite further constrains the form of the Ansatz studied. In Section
\ref{ymeqn}, we give the general form of the source free YM field equations.
We also examine in detail the existence of special trajectories for which
YM field equations are satisfied. We then give the series expansion of these
equations in powers of $1/R$. In Section \ref{charge}, the total color
charge is defined and its form is derived for the $1/R$ expansion of the
YM connection Ansatz under study. In Section \ref{sols}, we present two
new exact solutions as well as one approximate solution that is obtained
by solving the $1/R$ expanded field equations under the finite total
color charge constraint. In Section \ref{gauge}, we examine whether the
solutions presented are gauge equivalent to the Trautman solution \cite{traut}.
In two Appendices, we give the explicit form of some complicated equations
that are needed in the text.

\section{\label{curve} The Geometry of a Curve}

Let $z^{\mu} (\tau)$ define a smooth curve $C$ in a flat spacetime with Minkowski
metric $\eta_{\mu\nu}$\footnote{A detailed and careful analysis of the material
presented here can be found in \cite{gs1}. Here we only give a brief summary of
things that will be used and needed in this article.}. One can, in general,
define two times using an arbitrary point $x^{\mu}$ outside the curve.
Now let $\tau$ denote {\it the retarded time}
in the usual sense -- that one obtains by looking at the roots of
\( (x^{\mu} - z^{\mu} (\tau)) \, (x_{\mu} - z_{\mu} (\tau)) \, = 0\) -- and
define the null vector
\begin{equation}
\lambda_{\mu} \, \equiv \, \frac{\partial \, \tau}{\partial \, \mu} \,
= \, \frac{x_{\mu}-z_{\mu}(\tau)}{R},
\label{lam}
\end{equation}
where \( R \equiv \dot{z}^{\mu}(\tau) \, (x_{\mu}-z_{\mu}(\tau)) \) gives
{\it the retarded distance}. Here and from now on a dot over a letter denotes
differentiation with respect to the retarded time $\tau$. Differentiating
$\lambda_{\mu}$ and $R$, one finds that
\begin{eqnarray}
\lambda_{\mu, \nu} & = & \frac{1}{R} \, [\eta_{\mu \nu} - \dot{z}_{\mu} \,
\lambda_{\nu} - \dot{z}_{\nu} \, \lambda_{\mu} - (R \, a - \epsilon) \,
\lambda_{\mu} \lambda_{\nu}] \, , \\
R_{,\mu} & = & (R \, a - \epsilon) \, \lambda_{\mu} + \dot{z}_{\mu} \label{rder}
\end{eqnarray}
where {\it the acceleration} of the particle is \( a \equiv \frac{1}{R} \,
\ddot{z}^{\mu} \, (x_{\mu}-z_{\mu}(\tau)) \) and
\( \dot{z}^{\mu} \, \dot{z}_{\mu} = \epsilon \equiv 0, \pm 1 .\) [We choose
$\epsilon =-1$ for time-like curves and $\epsilon =0$ for null curves.] Moreover,
one also finds that \( \lambda^{\mu} \, \dot{z}_{\mu} = 1 \,,
\lambda^{\mu} \, R_{,\mu} = 1 \) and \(\lambda^{\mu} \, a_{,\mu} = 0 .\)

In fact, one notices that one can define\footnote{Please refer to \cite{gs1}
again for a detailed discussion of $a_{k}$.}
\[ a_{k} \equiv \lambda_{\mu} \, \frac{d^k \, \ddot{z}^{\mu}}{d \tau^k},
~~~k=1,2,\cdots ,D-1 \] which satisfy
\begin{equation}
\lambda^{\mu} \, a_{k,\mu} = 0 \, , \label{akmlam}
\end{equation}
for all $k$. Here $a_0 \equiv a$ and $D$ denotes the dimension of the spacetime.
[From now on we will take $D=4$.] The scalars $a_{k}$, which are
generalizations of the acceleration $a$ of the point particle, are related
to the curvature scalars of the smooth curve $C$ \cite{spi}, \cite{gs1}.

By using the curve $C$ and its kinematics, one can construct more general
solutions to the classical source free YM field equations than the one
found by Trautman \cite{traut}.

\section{\label{enflux} The Total Energy Flux}

Making use of the curve kinematics that has been briefly discussed in
Section \ref{curve}, we now make the following Ansatz for the LW potential
\begin{equation}
A_{\mu} = H \, \dot{z}_{\mu} + G \, \lambda_{\mu} \, , \label{vacans}
\end{equation}
where $H$ and $G$ are differentiable functions of $R$ and some $R$-independent
functions $c_i \; (i=1,2, \cdots)$ such that $\lambda^{\mu} \, c_{i,\mu} = 0$
for all $i$\footnote{Here we suppress the internal group indices on $A_{\mu}$.}. It
is clear that due to the property (\ref{akmlam}) of $a_k$, all of these
functions $(c_i)$ are functions of the scalars $a$ and $a_k \; (k=1,2,3)$
and the retarded time $\tau$. For the original Trautman solution, $G=0$
and $H=q/R$ where $q=q(\tau)$ only \cite{traut}. We are motivated to
choose the more general Ansatz above (\ref{vacans}) by our recent results
concerning the $D$-dimensional Einstein Maxwell theory with a null perfect
fluid in the Kerr-Schild geometry \cite{gs1}.

In close analogy to how it is defined in the Maxwell case
\cite{bar}, {\it the total energy flux} of the YM field is
given by\footnote{Please refer to \cite{gs1} for details.}
\begin{equation}
N_{YM} = - \int_{S} \, \dot{z}_{\mu} \, T^{\mu \nu} \, n_{\nu} \, R \, d\Omega \, .
\label{flux}
\end{equation}
Here $T_{\mu \nu} = F_{\mu \alpha}\, F_{\nu}\,^{\alpha} - \frac{1}{4} \,
\eta_{\mu \nu} \, F_{\alpha \beta} \, F^{\alpha \beta}$ is the YM energy
momentum tensor (with again the trace over the internal group indices
suppressed on the YM field strength $F$.) The vector $n_{\nu}$ is orthogonal
to the velocity vector field $\dot{z}_{\mu}$ and is defined through
\begin{equation}
\lambda_{\mu} = \epsilon \dot{z}_{\mu} + \epsilon_{1} \, \frac{1}{R} \, n_{\mu}
\;\;\;\; ; \; n^{\mu} \, n_{\mu} = - \epsilon R^2 \, . \label{lamzn}
\end{equation}
(Here $\epsilon_{1}= \pm 1$.) One can consider $S$ in the rest frame of the
point particle as a sphere $S^2$ of very large radius $R$. $d\Omega$, of course,
denotes the solid angle.

Substituting for $A_{\mu}$ (\ref{vacans}) in the YM field strength\footnote{In
this work we set the gauge-field coupling constant equal to one.}
\begin{equation}
F_{\mu\nu} = \partial_{\mu} A_{\nu} - \partial_{\nu} A_{\mu}
+ [A_{\mu}, A_{\nu}] \; , \label{fmn}
\end{equation}
one finds that
\[ F_{\mu\nu} = \lambda_{\mu} \chi_{\nu} - \lambda_{\nu} \chi_{\mu}
+ H_{,\mu} \dot{z}_{\nu} - H_{,\nu} \dot{z}_{\mu}\]
where for convenience we have defined
\( \chi_{\nu} \equiv H \ddot{z}_{\nu} - G_{,\nu} - [H,G] \dot{z}_{\nu} .\)

Using the orthogonality of the velocity vector field $\dot{z}_{\mu}$ and
the vector $n_{\mu}$, one finds that
\[ \dot{z}_{\mu} \, T^{\mu \nu} \, n_{\nu} =
(F_{\mu \alpha}\, \dot{z}^{\mu}) \, (F_{\nu}\,^{\alpha}\, n^{\nu}) \, ,\] or
substituting for $F_{\mu\nu}$ in this expression, that
\[ R\, \dot{z}_{\mu} \, T^{\mu \nu} \, n_{\nu} =
- \epsilon\, \epsilon_{1}\, R^2\, \mbox{Tr} \, (Y^{\alpha}\, J_{\alpha})\]
where
\begin{eqnarray*}
Y_{\alpha} & \equiv & H \ddot{z}_{\alpha} +
\dot{z}_{\alpha} \, (H_{,\mu} \dot{z}^{\mu} - \, [H,G]) \\
& & + \lambda_{\alpha} \, (G_{,\mu} \dot{z}^{\mu} + \epsilon \, [H,G]) -
G_{,\alpha} - \epsilon \, H_{,\alpha} \;\;\;\;\;\; \mbox{and} \\
J_{\alpha} & \equiv & H \ddot{z}_{\alpha} +
\dot{z}_{\alpha} \, (H_{,\mu} \dot{z}^{\mu} -[H,G] - \epsilon \, H^{\prime}) \\
& & + \lambda_{\alpha} \, (G_{,\mu} \dot{z}^{\mu} + \epsilon \, (a \, H - G^{\prime})) -
G_{,\alpha} \; .
\end{eqnarray*}
Here and from now on a prime over a letter denotes partial differentiation with
respect to $R$.

Notice that when \(G=G(R,c_i (\tau,a,a_{k})) \;\; (k=1,2,3)\) in its full generality,
\begin{eqnarray*}
G_{,\alpha} & = & G^{\prime} R_{,\alpha} + G_{,c_i} c_{i,\alpha} \\
& = & G^{\prime} R_{,\alpha} + G_{,c_i} (\dot{c_i} \lambda_{,\alpha} +
c_{i,a} a_{,\alpha} + c_{i,a_{k}} a_{k,\alpha} )
\end{eqnarray*}
and similarly for \(H=H(R,c_i (\tau,a,a_{k})) \;\; (k=1,2,3) \). The derivative
of the acceleration $a$ is
\begin{equation}
a_{,\alpha} = \frac{1}{R} \ddot{z}_{\alpha} - \frac{a}{R} \dot{z}_{\alpha}
+ (a_{1} - a^2 + \epsilon \, \frac{a}{R}) \lambda_{\alpha} \label{asubm}
\end{equation}
and similarly one can also calculate the derivatives of $a_{k}$ for $k=1,2,3$.
As an example, the derivative of $a_1$ is given in the Appendix \ref{app1},
(\ref{a1der}). It is easy to see that putting the derivatives of $a_{k}$, one
ends up with higher order $\tau$ derivatives of $z_{\mu}$ and more complicated expressions
for $G_{,\alpha}$ (and likewise for $H_{,\alpha}$) and hence $N_{YM}$. So to keep
the angular integrations that appear in the expression of $N_{YM}$
and also the calculations that one encounters in the solution of the YM
field equations simple, from now on we assume that the functions $c_{i}$, whose
properties were described above, are just functions of $\tau$ and $a$. Hence
one has \(G=G(R,c_i (\tau,a)) \) and similarly \(H=H(R,c_i (\tau,a)) \)
from now on. With these assumptions, substituting the derivative of $a$ and
the derivative of the retarded distance $R$ (\ref{rder}) in the relevant
expressions above, one finds that
$Y_{\alpha}$ and $J_{\alpha}$ can, respectively, be put in the form
\begin{eqnarray*}
Y_{\alpha} & = & \ddot{z}_{\alpha} Y_2 + \dot{z}_{\alpha} Y_1 + \lambda_{\alpha} Y_0 \\
J_{\alpha} & = & \ddot{z}_{\alpha} J_2 + \dot{z}_{\alpha} J_1 + \lambda_{\alpha} J_0 \; ,
\end{eqnarray*}
with six coefficients $Y_2, J_2, Y_1, J_1, Y_0, J_0$ to be determined.
Notice that this yields
\begin{eqnarray*}
\mbox{Tr} \, (Y^{\alpha} J_{\alpha}) & = & \ddot{z}^{\alpha} \ddot{z}_{\alpha} \,
\mbox{Tr} \, (Y_2 J_2) +
a \, \mbox{Tr} \, (Y_2 J_0 + Y_0 J_2) \\
& & + \, \epsilon \, \mbox{Tr} \, (Y_1 J_1) + \mbox{Tr} \, (Y_1 J_0 + Y_0 J_1) \;
\end{eqnarray*}
using the properties of the velocity vector field $\dot{z}_{\mu}$ and $\lambda_{\mu}$.

Notice at this point that it is still very difficult to work out the general form
of the energy flux formula $N_{YM}$ let alone to find solutions of the YM
field equations in this general setting. However demanding that $N_{YM}$ be finite
at very large values of $R$, one can in general assume that a series expansion
of $H$ and $G$ can be made in powers of $1/R$ (with $R \neq 0$, of course) as
\begin{eqnarray*}
H & \equiv & \alpha + \frac{\beta}{R} + \frac{\gamma}{R^2} + O(R^{-3}) \;\;\; \mbox{and}
\;\;\; \\
G & \equiv & \sigma + \frac{\omega}{R} + \frac{\delta}{R^2} + O(R^{-3}) \; .
\end{eqnarray*}
Here $\alpha, \beta, \gamma$ and $\sigma, \omega, \delta$ are the
functions $c_{i}$ that we have described above and are just functions of the
retarded time $\tau$ and the acceleration $a$.

With these assumptions then, one finds that
\begin{eqnarray*}
J_2 & \equiv & H - \frac{1}{R} G_{,c_i} \, c_{i,a} \\
& = & \alpha + \frac{1}{R} (\beta - \sigma_{,a})
+ \frac{1}{R^2} (\gamma - \omega_{,a}) + O(R^{-3}) \;\;\; \mbox{and} \\
Y_2 & \equiv & H - \frac{1}{R} (G_{,c_i} + \epsilon H_{,c_i}) \, c_{i,a} \\
& = & \alpha + \frac{1}{R} (\beta - \sigma_{,a} - \epsilon \, \alpha_{,a})
+ \frac{1}{R^2} (\gamma - \omega_{,a} - \epsilon \beta_{,a}) \\
& & + O(R^{-3}) \; ,
\end{eqnarray*}
which in turn implies that $\mbox{Tr} \, (Y_2 J_2) = \mbox{Tr} \, (\alpha^2)
+ O(1/R)$ and one has to set
$\alpha =0$ to get a finite energy flux as one takes the limit $R \to \infty$
in the expression (\ref{flux}) for $N_{YM}$. So one is now left with
\begin{eqnarray}
H & \equiv & \frac{\beta}{R} + \frac{\gamma}{R^2} + O(R^{-3}) \label{hser}
\;\;\; \mbox{and} \;\;\; \\
G & \equiv & \sigma + \frac{\omega}{R} + \frac{\delta}{R^2} + O(R^{-3})
\; , \label{gser}
\end{eqnarray}
where the five remaining coefficients $\beta, \gamma, \sigma, \omega$
and $\delta$ (functions of $\tau$ and $a$) are to be determined by the YM
field equations.

Carrying out the calculations of the remaining coefficients $Y_1, J_1, Y_0$
and $J_0$ in the same manner to $O(R^{-3})$ and using these, one finds that
\begin{eqnarray*}
\mbox{Tr} \, (Y_2 J_2) & = & \frac{1}{R^2} \mbox{Tr} \,
(\beta - \sigma_{,a})^2 + O(R^{-3}) \; , \\
a \, \mbox{Tr} \, (Y_2 J_0) & = & \epsilon \, \frac{a^2}{R^2}
\mbox{Tr} \, (\beta - \sigma_{,a})^2 + O(R^{-3}) \; ,\\
a \, \mbox{Tr} \, (Y_0 J_2) & = & - \mbox{Tr} \, (Y_1 J_0) \\
& = & \epsilon \frac{a}{R^2} \, \mbox{Tr} \,
\lbrace ([\beta,\sigma] + a(\beta - \sigma_{,a}) - \dot{\beta} \\
& &- (a_1 - a^2) \beta_{,a}) \times \, (\beta - \sigma_{,a}) \rbrace
+ O(R^{-3}) \; , \\
\epsilon \, \mbox{Tr} \, (Y_1 J_1) & = & - \mbox{Tr} \, (Y_0 J_1) \\
& = & \frac{\epsilon}{R^2} \mbox{Tr} \,
\lbrace [\beta,\sigma] + a(\beta - \sigma_{,a}) - \dot{\beta} \\
& & - (a_1 - a^2) \beta_{,a} \rbrace^2 + O(R^{-3}) \; ,
\end{eqnarray*}
and hence
\[ R\, \dot{z}_{\mu} \, T^{\mu \nu} \, n_{\nu} =
- \epsilon\, \epsilon_{1}\, (\ddot{z}^{\alpha} \ddot{z}_{\alpha} + \epsilon a^2) \,
\mbox{Tr} \, (\beta - \sigma_{,a})^2 + O(1/R) \; . \]

So taking the $R \to \infty$ limit, one finds
\begin{equation}
N_{YM} = \int_{S} \, d\Omega \, \epsilon \, \epsilon_{1} \,
(\ddot{z}^{\alpha} \ddot{z}_{\alpha} + \epsilon a^2) \,
\mbox{Tr} \, (\beta - \sigma_{,a})^2 \label{fluxba}
\end{equation}
for the energy flux.

Notice that the YM field equations have not been solved
yet and at this stage $N_{YM}$ contains only the first terms that appear in
the series expansion of $H$ and $G$ in powers of $1/R$ (\ref{hser}),
(\ref{gser}). So to observe any outgoing radiation at large distances,
one has to keep either $\beta$, the coefficient of the $1/R$ term in $H$,
and/or $\sigma$, the constant term in $G$ of the YM connection (\ref{vacans}).

We will come back to the discussion of the energy flux $N_{YM}$ after we
find solutions of the YM field equations using the YM connection (\ref{vacans})
with $H$ and $G$ given by (\ref{hser}) and (\ref{gser}).

\section{\label{ymeqn} The Source Free YM Equations}

The source free YM field equations simply read
\begin{equation}
D^{\mu} \, F_{\mu\nu} \, = \, \partial^{\mu} \, F_{\mu\nu} \, +
\, [A^{\mu},F_{\mu\nu}] \, = \, 0 \; .
\label{ymeq}
\end{equation}
The field strength is given by (\ref{fmn}), of course. Taking $A_{\mu}$
as in (\ref{vacans}) with \(H=H(R,c_i (\tau,a)) \) and \(G=G(R,c_i (\tau,a)) \),
calculating $F_{\mu\nu}$ and using these in (\ref{ymeq}), one finds that in the
general case $D^{\mu} F_{\mu\nu}$ can be put in the form
\begin{equation}
D^{\mu} \, F_{\mu\nu} = X \, z^{(3)}_{\nu} + Y \, \ddot{z}_{\nu}
+ K \, \dot{z}_{\nu} + L \, \lambda_{\nu} \label{xykl} \, = \, 0
\end{equation}
where $z^{(3)}_{\nu}$ denotes $d^{3} z_{\nu} / d \tau^3$. The explicit form of
$D^{\mu} F_{\mu\nu}$ is given in the Appendix \ref{app1}. One immediately
recognizes that it is extremely difficult to find exact solutions to
$D^{\mu} F_{\mu\nu} = 0$ (\ref{ymeq}) in this general setting. Even though
this is the case, in what follows we are going to try to consider every
possible case in detail, do what can be said and done and leave out only
the most complicated equations which we found too hard to solve.

\subsection{\label{traj} A Closer Look at the Trajectories}

Notice at this point that one can ask whether $D^{\mu} F_{\mu\nu} = 0$ is
satisfied identically without imposing any conditions on $H$ and/or $G$ in
$A_{\mu}$; i.e. whether there are any special trajectories that a point
particle can follow so that the YM equations identically hold outside its
position.

Contracting $D^{\mu} F_{\mu\nu}$ with $\lambda^{\nu}, \dot{z}^{\nu}, \ddot{z}^{\nu},$
etc. one finds that $D^{\mu} F_{\mu\nu}$ can in general be written as
\[ D^{\mu} F_{\mu\nu} = X (z^{(3)}_{\nu}
+ p \ddot{z}_{\nu} + q \dot{z}_{\nu} + r \lambda_{\nu}) \]
where
\begin{eqnarray*}
p & \equiv & \frac{1}{\Delta} (\ddot{z}^{\mu} \, z^{(3)}_{\mu} + a (\epsilon \, a_{1}
+ \ddot{z}^{\mu} \, \ddot{z}_{\mu})) \; , \\
q & \equiv & - \frac{1}{\Delta} (a \, \ddot{z}^{\mu} \, z^{(3)}_{\mu}
+ \ddot{z}^{\mu} \, \ddot{z}_{\mu} \, (a^{2}- a_{1})) \; , \\
r & \equiv & \frac{1}{\Delta} (\epsilon \, a \, \ddot{z}^{\mu} \, z^{(3)}_{\mu}
- (\epsilon \, a_{1} + \ddot{z}^{\mu} \, \ddot{z}_{\mu}) \, \ddot{z}^{\alpha} \, \ddot{z}_{\alpha}) \; , \\
\Delta & \equiv & - (\epsilon \, a^2 + \ddot{z}^{\mu} \, \ddot{z}_{\mu}) \neq 0 \; .
\end{eqnarray*}

Now setting $X=0$ would imply that $H_{,c_{i}} c_{i,a} = 0$. Since $H_{,c_{i}}=0$
necessarily excludes the curve kinematics that we want to introduce into the picture,
this leaves one with $c_{i,a} =0$ or $c_{i} = c_{i}(\tau)$. [Notice that $G$ still
may depend on $c_{i}$ with $c_{i}=c_{i}(\tau,a)$.] However in that case
$D^{\mu} F_{\mu\nu}$ now takes the form
\[ D^{\mu} F_{\mu\nu} = M_{2} \ddot{z}_{\nu} + M_{1} \dot{z}_{\nu} + M_{0} \lambda_{\nu} \]
where, for example,
\[ M_2 = H^{\prime} + \frac{H}{R} - \frac{1}{R}
(G^{\prime}_{,c_{i}} + [H,G_{,c_{i}}]) c_{i,a} \; . \]
One is then either forced to take the trajectory curve $C$ as a straight line or
set $M_{2}=M_{1}=M_{0}=0$ for a nontrivial curve $C$. However these three equations
$M_{2}=M_{1}=M_{0}=0$ are again highly nonlinear and very complicated to work with.

For $X \neq 0$, one has to put
\begin{equation}
z^{(3)}_{\nu} + p \ddot{z}_{\nu} + q \dot{z}_{\nu} + r \lambda_{\nu} = 0 \; .
\label{cureq}
\end{equation}
However making use of the Serret-Frenet frame in four dimensions \cite{gs1}, one finds
that
\[ p = - \frac{\dot{\kappa}_{1}}{\kappa_{1}} \;\; , \;\; q = - \kappa_{1}^2
\;\; \mbox{and} \;\; r=0 \]
and that (\ref{cureq}) is satisfied identically when $i) \, \kappa_{1}=0$, i.e. $a=0$
which implies that the trajectory curve $C$ is a straight line or $ii) \,$
$\kappa_{2}=0$ but $\kappa_{1} \neq 0$. The first case again gives a
``trivial" solution, whereas the second case implies that one has to both constrain
the trajectory curve $C$ and find the corresponding $H$ and $G$ which satisfy
\[ Y = - \frac{\dot{\kappa}_{1}}{\kappa_{1}} X \;\; , \;\; K = - \kappa_{1}^2 X
\;\; \mbox{and} \;\; L=0 \; .\]
These equations are again very difficult to solve. So instead of working out
these complicated conditions on the trajectory curve $C$, we now concentrate on
what one can do with the YM equations themselves.

\subsection{\label{asymym} $1/R$ Expansion of Source Free YM Equations}

Since it is very hard to find exact solutions to
$D^{\mu} F_{\mu\nu} = 0$ (\ref{xykl}) in its full generality, we look
for approximate solutions by using the series expansion of $H$ and $G$
in powers of $1/R$ (\ref{hser}), (\ref{gser}). Substituting these
in the expressions for $X, Y, K$ and $L$ (see equations (\ref{xeq}),
(\ref{yeq}), (\ref{keq}) and (\ref{leq})), one finds to order $R^{-3}$ that
\begin{eqnarray*}
D^{\mu} F_{\mu\nu} & = & \frac{1}{R} \lambda_{\nu} (L_{0}) \\
& & + \frac{1}{R^2} (X_{1} \, z^{(3)}_{\nu} + Y_{1} \, \ddot{z}_{\nu}
+ K_{1} \, \dot{z}_{\nu} + L_{1} \, \lambda_{\nu}) \\
& & + O(R^{-3}) \, ,
\end{eqnarray*}
where
\begin{widetext}
\begin{eqnarray}
L_{0} & \equiv & a \{ \dot{\beta} +  (a_{1} - a^2) \beta_{,a} - [\beta,\sigma] \}
+ 2 [\dot{\beta} +  (a_{1} - a^2) \beta_{,a}, \sigma]
+ [\beta, \dot{\sigma} +  (a_{1} - a^2) \sigma_{,a}] + [\sigma,[\beta,\sigma]]
\nonumber \\
& & - \beta_{,a} (a_{2} - 3 a_{1} a + 2 a^{3})
- 2 \, (a_{1} - a^2) \, \dot{\beta}_{,a} - (a_{1} - a^2)^2 \, \beta_{,aa}
- \ddot{\beta} \; , \label{l0eq} \\
L_{1} & \equiv & 2 a \omega - \dot{\omega} - (a_{1} - a^2) \, \omega_{,a}
+ 2 \epsilon \, a \sigma_{,a} + (\epsilon \, a^2 + \ddot{z}^{\alpha} \, \ddot{z}_{\alpha}) \,
\sigma_{,aa} + (a_{1} - 3 a^2) \gamma - \epsilon \, \dot{\beta}
- \epsilon \, a \dot{\beta}_{,a} \nonumber \\
& & - \epsilon (2 a_{1} - 3 a^2 + \epsilon \, \ddot{z}^{\alpha} \, \ddot{z}_{\alpha}) \beta_{,a}
- \epsilon \, a (a_{1} - a^2) \beta_{,aa} - (a_{1} - a^2)^{2} \, \gamma_{,aa}
- 2 (a_{1} - a^2) \dot{\gamma}_{,a} - \ddot{\gamma} + 3 a \dot{\gamma}
\nonumber \\
& & - (a_{2} - 6 a_{1} a + 5 a^3) \, \gamma_{,a} + \epsilon \, [\beta,\sigma]
- 3 a [\gamma,\sigma]
- 2 a [\beta,\omega] + 2 [\dot{\beta} +  (a_{1} - a^2) \, \beta_{,a},\omega]
\nonumber \\
& & + 2 [\dot{\gamma} +  (a_{1} - a^2) \, \gamma_{,a},\sigma]
+ [\gamma,\dot{\sigma} +  (a_{1} - a^2) \, \sigma_{,a}]
+ [\beta,\dot{\omega} +  (a_{1} - a^2) \, \omega_{,a}]
+ \epsilon \, [\beta,[\beta,\sigma]] \nonumber \\
& & - [\sigma,\omega] + [\sigma,[\beta,\omega]]
+ [\omega,[\beta,\sigma]] + [\sigma,[\gamma,\sigma]]
- \epsilon \, [\beta,\dot{\beta} +  (a_{1} - a^2) \, \beta_{,a}]
- \epsilon \, a [\sigma,\beta_{,a}] - \epsilon \, a [\beta,\sigma_{,a}] \; , \label{l1eq} \\
K_{1} & \equiv & \dot{\beta} + (2 a_{1} - 3 a^2) \beta_{,a}
+ a (\dot{\beta}_{,a} + (a_{1} - a^2) \, \beta_{,aa}) + [\sigma,\beta] + [\beta,\dot{\beta}]
+ (a_{1} - a^2) \, [\beta,\beta_{,a}] - [\beta,[\beta,\sigma]] \nonumber \\
& & + a ([\sigma,\beta_{,a}] + [\beta,\sigma_{,a}]) \; , \label{k1eq} \\
Y_{1} & \equiv & - [\sigma,\beta_{,a}] - [\beta,\sigma_{,a}]
- \dot{\beta}_{,a} - (a_{1} - a^2) \, \beta_{,aa} + 2 \, a \beta_{,a} \; , \label{y1eq} \\
X_{1} & \equiv & - \beta_{,a} \; . \label{x1eq}
\end{eqnarray}
\end{widetext}

Before dwelling on the solutions of the equations above, we now make a digression
and briefly review the definition of a gauge-invariant total color charge. Demanding
the total color charge to be finite at large $R$ values constrains the system of differential
equations to be solved considerably and one is, at least, able to talk about the
behavior of solutions.

\section{\label{charge} Total Color Charge $Q$}

In the presence of sources, the YM field equations are
\begin{equation}
D^{\mu} \, F_{\mu\nu} \, = \, \partial^{\mu} \, F_{\mu\nu} \, +
\, [A^{\mu},F_{\mu\nu}] \, = \, J_{\nu} \; .
\label{symeq}
\end{equation}
Even though $D^{\mu} J_{\mu} = 0$, using this, one can not define a conserved or
a gauge-invariant charge in the usual sense. However, one can make use of the
fact that
\[ \partial^{\nu} \, \partial^{\mu} \, F_{\mu\nu} = \partial^{\nu} \,
(J_{\nu} - [A^{\mu},F_{\mu\nu}] ) \, = 0 \]
and define the total color as
\begin{equation}
I = \int \, d^3x \, (J_{0} - [A^{i},F_{i0}]) \, .
\label{ftocol}
\end{equation}
According to \cite{olt}, considering gauge transformations that are independent
of space-time coordinates at large distances; i.e. taking into account YM
connections $A_{\mu}$ that go to zero faster than $1/R$ at large $R$, $I$ turns
out to be gauge covariant and a gauge invariant {\it total color charge} can be defined
as\footnote{The internal group indices on $Q$ and $I$ are suppressed throughout.}
(see the discussions in \cite{tt} and \cite{olt} as well)
\begin{equation}
Q \, = \, \sqrt{\mbox{Tr} \, I^2} \; .
\label{ch}
\end{equation}
However it is also claimed in \cite{lo}, \cite{olt} that the total color
(\ref{ftocol}) is not well suited for determining the color exchange between
an external source and the YM waves. The reason they give is that neither
(\ref{ftocol}) nor (\ref{ch}) can be written in a gauge independent manner
as a sum of the gauge invariant color of the external source and that of the
YM field.

Here we are not going to take part in this discussion. Whether it can
be written in a gauge independent manner or not, for (\ref{ch}) to make sense
as the definition of some charge, (\ref{ftocol}) must, first of all, be finite
when the integration is carried on the full space. This is especially essential
for us in this work, since the only source here is a point source moving along a
trajectory. However some attention is also needed here: One has to do the
integration on such a surface that it respects the motion of the point source.

Notice that similar considerations led \cite{bar} to the expression for the
total energy flux $N_{YM}$ in (\ref{flux}). So in complete analogy, one can
define the total color as
\begin{equation}
I = \int_{S} \, d\Omega \, R^2 \, (J_{\nu} - [A^{\mu},F_{\mu\nu}]) \, \dot{z}^{\nu} \; ,
\label{stocol}
\end{equation}
or simply as
\begin{equation}
I = \int_{S} \, d\Omega \, R^2 \, (\partial^{\mu} \, F_{\mu\nu}) \, \dot{z}^{\nu} \; ,
\label{tocol}
\end{equation}
so that the motion of the point source is taken into account. The surface $S$ can
again be thought of as a sphere $S^2$ of very large radius $R$ in the rest frame
of the point source. One can again use (\ref{ch}) as the definition of the total
color charge then.

So we need to examine $\partial^{\mu} \, F_{\mu\nu}$ for the $A_{\mu}$ Ansatz
(\ref{vacans}) that we have. Similar to what we did for $D^{\mu} F_{\mu\nu}$,
substituting $H$ and $G$ in powers of $1/R$ (\ref{hser}), (\ref{gser}) in the
expression for $\partial^{\mu} \, F_{\mu\nu}$ (see Appendix \ref{app2}), one
finds to order $R^{-3}$ that
\begin{eqnarray*}
\partial^{\mu} F_{\mu\nu} & = & \frac{1}{R} \lambda_{\nu} (S_{0}) \\
& & + \frac{1}{R^2} (B_{1} \, z^{(3)}_{\nu} + C_{1} \, \ddot{z}_{\nu}
+ P_{1} \, \dot{z}_{\nu} + S_{1} \, \lambda_{\nu}) \\
& & + O(R^{-3}) \, ,
\end{eqnarray*}
where
\begin{widetext}
\begin{eqnarray}
S_{0} & \equiv & a \{ \dot{\beta} +  (a_{1} - a^2) \beta_{,a} - [\beta,\sigma] \}
+ [\dot{\beta} +  (a_{1} - a^2) \beta_{,a}, \sigma]
+ [\beta, \dot{\sigma} +  (a_{1} - a^2) \sigma_{,a}]
- \beta_{,a} (a_{2} - 3 a_{1} a + 2 a^{3})
 \nonumber \\
& &  - 2 (a_{1} - a^2) \dot{\beta}_{,a} - (a_{1} - a^2)^2 \, \beta_{,aa}
- \ddot{\beta} \; , \label{s0eq} \\
S_{1} & \equiv & 2 a \omega - \dot{\omega} - (a_{1} - a^2) \, \omega_{,a}
+ 2 \epsilon \, a \sigma_{,a} + (\epsilon \, a^2 + \ddot{z}^{\alpha} \, \ddot{z}_{\alpha}) \,
\sigma_{,aa} + (a_{1} - 3 a^2) \gamma - \epsilon \, \dot{\beta}
- \epsilon \, a \dot{\beta}_{,a} \nonumber \\
& & - \epsilon (2 a_{1} - 3 a^2 + \epsilon \, \ddot{z}^{\alpha} \, \ddot{z}_{\alpha}) \beta_{,a}
- \epsilon \, a (a_{1} - a^2) \beta_{,aa} - (a_{1} - a^2)^{2} \, \gamma_{,aa}
- 2 (a_{1} - a^2) \dot{\gamma}_{,a} - \ddot{\gamma} + 3 a \dot{\gamma}
\nonumber \\
& & - (a_{2} - 6 a_{1} a + 5 a^3) \, \gamma_{,a} - 2 a [\gamma,\sigma]
- 2 a [\beta,\omega] + [\dot{\beta} +  (a_{1} - a^2) \, \beta_{,a},\omega]
+ [\dot{\gamma} +  (a_{1} - a^2) \, \gamma_{,a},\sigma] \nonumber \\
& & + [\beta,\dot{\omega} +  (a_{1} - a^2) \, \omega_{,a}]
+ [\gamma, \dot{\sigma} +  (a_{1} - a^2) \, \sigma_{,a}] \; , \label{s1eq} \\
P_{1} & \equiv & \dot{\beta} + (2 a_{1} - 3 a^2) \beta_{,a}
+ a (\dot{\beta}_{,a} + (a_{1} - a^2) \, \beta_{,aa}) + [\sigma,\beta] \; , \label{p1eq} \\
C_{1} & \equiv &  - \dot{\beta}_{,a} - (a_{1} - a^2) \, \beta_{,aa}
+ 2 \, a \beta_{,a} \; , \label{c1eq} \\
B_{1} & \equiv & - \beta_{,a} \; . \label{b1eq}
\end{eqnarray}
\end{widetext}

Since $\lambda^{\nu} \dot{z}_{\nu} =1$, one is forced to set $S_{0} = 0$ so that
$I$ in (\ref{tocol}) and hence $Q$ become finite for very large $R$. Hence taking
the $R \to \infty$ limit, the total color charge $Q$ is defined through $I$
in (\ref{tocol}) which turns out to be
\[ I = \int_{S} \, d\Omega \, (-B_{1} \, \ddot{z}^{\alpha} \, \ddot{z}_{\alpha}
+ \epsilon \, P_{1} + S_{1}) \]
for the special form of the $A_{\mu}$ Ansatz (\ref{vacans}) that we are using.

\section{\label{sols} Solutions}

In this Section we look for solutions to the source free YM equations. We first
start with the original Trautman solution to remind the reader about its
properties and also to check the calculations that have been done so far.
We then give two new exact solutions and finally present the ``approximate" one
which is obtained by using a series expansion in $1/R$.

\subsection{\label{trasol} Trautman Solution}

In the case when $G=0$, i.e. $\sigma=\omega=\delta=\gamma=0$ and
when $A_{\mu}$ is of the form
\begin{equation}
A_{\mu} \, = \, \frac{\beta}{R} \, \dot{z}_{\mu} \; ,
\label{solseca}
\end{equation}
one should of course find the original Trautman solution which is the first
example of a non-Abelian LW potential. Indeed, one finds in this case that
the YM field equations $D^{\mu} F_{\mu\nu}=0$ are satisfied {\it exactly}
provided $i) \, \beta = \beta(\tau)$, $ii) \, \ddot{\beta} - a \, \dot{\beta} = 0$
and $iii) \, \dot{\beta} + [\beta,\dot{\beta}]=0$,
as found by Trautman \cite{traut}. In this case since $\sigma=0$, the total
energy flux formula $N_{YM}$ is just a simple generalization of the
corresponding expression for the ordinary Abelian Maxwell theory, that is
obtained by replacing the square of the electric charge by $\mbox{Tr} \beta^2$
\cite{bar}, \cite{traut}, \cite{gs1}. Moreover, one also finds that
$S_{0}=0, S_{1}= - \epsilon \, \dot{\beta}, P_{1}= \dot{\beta}, B_{1}=C_{1}=0$
and hence $(\partial^{\mu} F_{\mu\nu}) \, \dot{z}^{\nu} =0$ identically in
this case. So the total color charge $Q=0$ and automatically conserved as
expected from \cite{traut}.

\subsection{\label{bg0d0} $\beta=\gamma=0 \, , \delta=0$}

For the special choice of $\beta=\gamma=0$ and $\delta=0$, i.e. when
$A_{\mu}$ is of the form
\begin{equation}
A_{\mu} \, = \, (\sigma + \frac{\omega}{R} ) \, \lambda_{\mu} \; ,
\label{solsecb}
\end{equation}
one finds that the YM field equations $D^{\mu} F_{\mu\nu}=0$ are satisfied
{\it exactly} provided that $\sigma$ and $\omega$ satisfy
\begin{eqnarray}
\omega & = & \omega(\tau) , ~~\label{oot} \\
\partial_{a} \, (\sigma_{,a} (\ddot{z}^{\alpha} \ddot{z}_{\alpha} + \epsilon \, a^2)
+ a^2 \, \omega) - \dot{\omega} - [\sigma,\omega] & = & 0 \, . ~~\label{omeq}
\end{eqnarray}
Notice that in this case the total energy flux $N_{YM}$  expression (\ref{fluxba})
has only $\beta - \sigma_{,a} = - \sigma_{,a}$ in it, which is considerably
different than the Trautman solution in character. The dependence of $\sigma$
on the acceleration $a$ of the trajectory turns out to determine the form of
$N_{YM}$. Moreover, in this case one finds that $B_{1}=C_{1}=P_{1}=0$ as well as
$S_{0}=0$, of course, but $S_{1}=[\sigma,\omega]$. Hence $I$ now becomes
\( I = \int_{S} \, d\Omega \, [\sigma,\omega]\). $Q=0$ iff $[\sigma,\omega]=0$,
of course.

A trivial solution to (\ref{omeq}) is given by \( \sigma_{,aa}=0 \, ,
\epsilon \, \sigma_{,a} + \omega =0 \) and $\dot{\omega} + [\sigma,\omega]=0$.
Then \(\sigma = a k(\tau) + l(\tau) \, , \omega = - \epsilon k(\tau)\) and
the arbitrary functions $k(\tau)$ and $l(\tau)$ satisfy $\dot{k} + [l,k]=0$.
(If one chooses $k=-q$ and $l=-\dot{q}$, then $\sigma = -a q(\tau) - \dot{q}(\tau)$
and $\omega = \epsilon \, q$, and this yields the same condition that one obtains
in the case of the Trautman solution. One, of course, expects that this trivial
solution is gauge equivalent to Trautman's original solution.) Notice also that
then \( I = \int_{S} \, d\Omega \, (- \epsilon \, [l,k]) =
\int_{S} \, d\Omega \, \epsilon \, \dot{k} \) and
$\beta - \sigma_{,a} = - \sigma_{,a} = - k(\tau)$ in $N_{YM}$ (\ref{fluxba}).

\subsection{\label{g0d0} $\gamma=0 \, , \delta=0$}

For the special choice of $\gamma=0$ and $\delta=0$, i.e. when
$A_{\mu}$ is of the form
\begin{equation}
A_{\mu} \, = \frac{\beta}{R} \, \dot{z}_{\mu}
+ (\sigma + \frac{\omega}{R} ) \, \lambda_{\mu} \; ,
\label{solsecc}
\end{equation}
one finds again that the YM field equations $D^{\mu} F_{\mu\nu}=0$ are
satisfied {\it exactly} provided that $\beta$, $\sigma$ and $\omega$ satisfy
\begin{eqnarray}
\beta & = & \beta(\tau) , ~~\label{gd0eq1} \\
\dot{\beta} - [\beta,\sigma] & = & 0 , ~~\label{gd0eq2} \\
\omega + g(\tau) - [\beta,\omega] & = & 0 , ~~\label{gd0eq3} \\
\lbrack \beta,g \rbrack & = & 0 , ~~\label{gd0eq4} \\
\dot{g} + \partial_{a} \, (\sigma_{,a} (\ddot{z}^{\alpha} \ddot{z}_{\alpha}
+ \epsilon \, a^2) - a^2 g) + [\sigma,g] & = & 0 , ~~\label{gd0eq5} \\
\partial_{a} \, (\omega_{,a} (\ddot{z}^{\alpha} \ddot{z}_{\alpha} + \epsilon \, a^2))
+ \epsilon \, (\omega + g) + [\omega,g] & = & 0 . ~~\label{gd0eq6}
\end{eqnarray}
Here $g=g(\tau)$ is an arbitrary function of $\tau$ that is obtained in one
of the integrations in the mid steps of the calculation. For this case one finds that $N_{YM}$
maintains its most general form and that $B_{1}=C_{1}=P_{1}=0$, as well as
$S_{0}=0$, of course, and $S_{1}= - \epsilon \, \dot{\beta} - [\sigma,g]$
which gives \( I = \int_{S} \, d\Omega \, (- \epsilon \, \dot{\beta} - [\sigma,g]) \).
Again $Q \neq 0$ for the general case and can be chosen to depend on $\tau$.

A trivial solution to equations (\ref{gd0eq5}) and (\ref{gd0eq6}) are given by choosing
\[ \sigma_{,aa}=0 \, ,
\epsilon \, \sigma_{,a} - g =0 \, , \dot{g} + [\sigma,g] =0 \, , \omega_{,a} =0 \, , \]
\[ \epsilon \, (\omega + g) + [\omega,g] =0 \, . \]
All of the conditions (\ref{gd0eq1}) to (\ref{gd0eq6}) are satisfied identically provided
\[ g(\tau) = \epsilon \, \beta \, , \sigma = a \beta + \sigma_{0}(\tau) \, ,
\omega = \omega(\tau) \, , \dot{\beta} + [\sigma_{0},\beta] =0 \]
and $[\beta,\omega]=\omega + \epsilon \, \beta$. (Here $\sigma_{0}(\tau)$ is
an arbitrary function and if one chooses $\sigma_{0} = - \dot{\beta}$ and
$\beta=q$, this yields the same condition that one finds for $q$ in the
Trautman solution.) Notice also that then $\beta - \sigma_{,a} = 0$ in $N_{YM}$
(\ref{fluxba}) and $I=0$, and hence $Q=0$.

\subsection{\label{gensol} General Case}

In this case we take $A_{\mu}$ to be of the form (\ref{vacans}) with
$H$ and $G$ given by (\ref{hser}) and (\ref{gser}), respectively.
We look for solutions of the YM field equations to order $R^{-3}$ by
setting $L_{0}=L_{1}=K_{1}=Y_{1}=X_{1}=0$ (\ref{l0eq}), (\ref{l1eq}), (\ref{k1eq}),
(\ref{y1eq}), (\ref{x1eq}). However due to the discussion at the end of Section
\ref{charge}, we also need to set $S_{0}=0$ (\ref{s0eq}) for a finite total
color charge. Hence we now have to solve these six equations simultaneously
for the five unknown coefficients $\beta, \gamma, \sigma, \omega$ and $\delta$
(which are, remember, only functions of $\tau$ and $a$).

In this case, one finds that $L_{0}=L_{1}=K_{1}=Y_{1}=X_{1}=0$ and $S_{0}=0$
are satisfied provided that
\begin{eqnarray}
\beta & = & \beta(\tau) , \label{geneq1} \\
\gamma & = & \gamma(\tau) , \label{geneq2} \\
\dot{\beta} - [\beta,\sigma] & = & 0 , \label{geneq3} \\
a \gamma - \omega - \tilde{g}(\tau) + [\beta,\omega] + [\gamma,\sigma] & = & 0 , \label{geneq4} \\
\partial_{a} \, (\sigma_{,a} (\ddot{z}^{\alpha} \ddot{z}_{\alpha}
+ \epsilon \, a^2) - a^2 (\tilde{g} - \dot{\gamma}))  & & \nonumber \\
+ \partial_{\tau} \, (\tilde{g} - \dot{\gamma})
+ [\sigma, \tilde{g} - \dot{\gamma}] & = & 0 . \label{geneq5}
\end{eqnarray}
Here $\tilde{g}=\tilde{g}(\tau)$ is an arbitrary function of $\tau$ that appears
in one of the integrations in the mid steps of the calculation. In this case one
finds that \( I = \int_{S} \, d\Omega \,
(- [\sigma, \tilde{g} - \dot{\gamma}]- \epsilon \, \dot{\beta})\)
and hence for the general case $Q \neq 0$ and, in fact, may be chosen to
depend on $\tau$. However notice that when one chooses the integration function
$\tilde{g}(\tau)$ as $\tilde{g}(\tau)=\epsilon \, \beta + \dot{\gamma}$, (\ref{geneq5})
becomes
\[ \sigma_{,a} (\ddot{z}^{\alpha} \ddot{z}_{\alpha} + \epsilon \, a^2)
- \epsilon \, \beta \, a^2 = n(\tau) \]
where $n(\tau)$ is a new arbitrary function of $\tau$, and $I=0$ and hence $Q=0$.
One can further choose $\sigma_{,a}=\beta$, i.e. $\sigma=a \beta(\tau) + \sigma_{0}(\tau)$,
and the arbitrary function $n(\tau)$ suitably such that this is also identically satisfied.
So now (\ref{geneq3}) implies $\dot{\beta} - [\beta,\sigma_{0}] = 0$ and a trivial solution
to (\ref{geneq4}) is provided by demanding that $i) \, \gamma + [\gamma,\beta]=0$ and
$ii) \, [\beta,\omega] + [\gamma,\sigma_{0}] - \omega - \tilde{g} =0$.

\section{\label{gauge} Gauge Equivalence}

One natural question to ask at this stage is, of course, whether the solutions
that have been found so far are gauge equivalent to the Trautman
solution \cite{traut}, also derived in Subsection \ref{trasol}. To answer this
question, one has to examine whether there exist any gauge potentials $\Phi$
(we again suppress internal group indices on $\Phi$) which locally satisfy
\begin{equation}
A_{\mu}^{Trautman} \, = \, A_{\mu}^{new sol} \, + \partial_{\mu} \, \Phi +
[A_{\mu}^{new sol},\Phi] \, = \, \frac{q}{R} \, \dot{z}_{\mu} \, . \label{gau1}
\end{equation}
(Here we take the original form of the Trautman solution, i.e. one has $q$
in place of $\beta$.)

Notice that substituting the general form of our Ansatz
$A_{\mu}=H \dot{z}_{\mu} + G \lambda_{\mu}$ (\ref{vacans}) and solving
for $\partial_{\mu} \Phi$, one finds that in general $\partial_{\mu} \Phi$
is of the form
\begin{equation}
\partial_{\mu} \, \Phi \, = \, X \, \dot{z}_{\mu} + Y \, \lambda_{\mu} \, .
\label{gau2}
\end{equation}
Demanding that $\Phi$ has continuous second order derivatives and that
$\partial_{\mu} \partial_{\nu} \Phi = \partial_{\nu} \partial_{\mu} \Phi$,
one finds
\begin{equation}
X_{,\nu} \dot{z}_{\mu} - X_{,\mu} \dot{z}_{\nu}
+ X (\ddot{z}_{\mu} \lambda_{\nu} - \ddot{z}_{\nu} \lambda_{\mu})
+ Y_{,\nu} \lambda_{\mu} - Y_{,\mu} \lambda_{\nu} \, = \, 0 \, .
\label{gau3}
\end{equation}
Contracting this with $\lambda^{\mu}$ and $\dot{z}^{\mu}$, one obtains two
equations which then can be solved for $X_{,\nu}$ and $Y_{,\nu}$ to yield
\begin{eqnarray}
X_{,\nu} & = & \dot{z}_{\nu} (\lambda^{\mu} X_{,\mu})
+ \lambda_{\nu} (\lambda^{\mu} Y_{,\mu} - a \, X) \, , \label{xnueq} \\
Y_{,\nu} & = & \ddot{z}_{\nu} X + \dot{z}_{\nu}
(\dot{z}^{\mu} X_{,\mu} - \epsilon \, \lambda^{\mu} X_{,\mu}) \nonumber \\
& & + \lambda_{\nu} (\dot{z}^{\mu} Y_{,\mu} - \epsilon \, \lambda^{\mu} Y_{,\mu}
- \epsilon \, a \, X) \, . \label{ynueq}
\end{eqnarray}
Substituting these into (\ref{gau3}), one finally finds that
\[ (\lambda^{\mu} Y_{,\mu} - a \, X - \dot{z}^{\mu} X_{,\mu}
+ \epsilon \, \lambda^{\mu} X_{,\mu})
(\dot{z}_{\mu} \lambda_{\nu} - \dot{z}_{\nu} \lambda_{\mu}) = 0 \, . \]
For a nontrivial gauge potential $\Phi$, one has to demand that
\begin{equation}
\lambda^{\mu} Y_{,\mu} - a \, X - \dot{z}^{\mu} X_{,\mu}
+ \epsilon \, \lambda^{\mu} X_{,\mu} = 0 \label{omcon}
\end{equation}
for the coefficients $X$ and $Y$ in (\ref{gau2}).

So now we look for the existence of such gauge potentials for each of the
solutions presented in Section \ref{sols} in order of their appearance.

\subsection{\label{gbg0d0} $\beta=\gamma=0 \, , \delta=0$}

For this {\it exact} solution, $X$ and $Y$ in (\ref{gau2}) turns out to be
\[ X = \frac{q}{R} \;\; , \;\; Y = - (\sigma + \frac{\omega}{R})
- [\sigma + \frac{\omega}{R},\Phi] \]
and remember that $q=q(\tau), \sigma=\sigma(\tau,a)$ and $\omega=\omega(\tau)$,
and these satisfy (\ref{omeq}) in this case. So imposing (\ref{omcon}), one gets
\[ \frac{1}{R} (- \dot{q} - [\sigma,q])
+ \frac{1}{R^2} (\omega - \epsilon \, q + [\omega,\Phi] - [\omega,q]) = 0 \, . \]
Since the coefficients $q,\sigma$ and $\omega$ are $R$ independent, one has to set
\begin{eqnarray}
\dot{q} + [\sigma,q] & = & 0 \; , \label{con1} \\
\omega - \epsilon \, q + [\omega,\Phi] - [\omega,q] & = & 0 \; . \label{con2}
\end{eqnarray}

A trivial solution is provided by $\sigma = - a q - \dot{q}, \omega= \epsilon \, q$
(see the end of Subsection \ref{bg0d0}) and $\Phi=k q$, where $k$ is an arbitrary
real number. In that case (\ref{oot}) and (\ref{omeq}) are also identically
satisfied since $q$, now being part of the Trautman solution, obeys
$\dot{q} + [q,\dot{q}]=0$. In this case \( A_{\mu}^{new sol} =
(-\dot{q} - a \, q + \epsilon \, \frac{q}{R}) \lambda_{\mu} \) and
this is gauge equivalent to the Trautman solution.

However in the general case when $\sigma=\sigma(\tau,a)$, it is not easy to find
a simultaneous solution to (\ref{oot}), (\ref{omeq}), (\ref{con1}) and (\ref{con2}),
and hence this class of exact solutions is not necessarily gauge equivalent to the
Trautman solution.

\subsection{\label{gg0d0} $\gamma=0 \, , \delta=0$}

For this {\it exact} solution, following similar steps as in Subsection \ref{gbg0d0}, one
finds that
\[ X = \frac{q-\beta}{R} - [\frac{\beta}{R},\Phi] \;\; , \;\;
Y = - (\sigma + \frac{\omega}{R}) - [\sigma + \frac{\omega}{R},\Phi] \]
and for this case remember that $\beta=\beta(\tau), \sigma=\sigma(\tau,a)$ and $\omega=\omega(\tau,a)$, and these satisfy (\ref{gd0eq1}), (\ref{gd0eq2}),
(\ref{gd0eq3}), (\ref{gd0eq4}), (\ref{gd0eq5}) and (\ref{gd0eq6}). Imposing
(\ref{omcon}), one finds that
\begin{eqnarray*}
\frac{1}{R} \lbrace \dot{\beta} - \dot{q} - [\sigma,q] + [\sigma,[\beta,\Phi]]
+ [\dot{\beta},\Phi] - [\beta,[\sigma,\Phi]] \rbrace \\
+ \frac{1}{R^2} \lbrace \omega - \epsilon \, (q-\beta) - [\omega,q]
+ [\omega + \epsilon \, \beta,\Phi] \\
+ [\omega,[\beta,\Phi]]
- [\beta,[\omega,\Phi]] \rbrace & = & 0 \, .
\end{eqnarray*}
Since the coefficients $\beta, \sigma, \omega$ and $q$ are $R$ independent,
one has to set
\begin{eqnarray}
\dot{\beta} - \dot{q} - [\sigma,q] + [\sigma,[\beta,\Phi]]
+ [\dot{\beta},\Phi] - [\beta,[\sigma,\Phi]] & = & 0 \, , ~~~ \label{cond1} \\
\omega - \epsilon \, (q-\beta) - [\omega,q] + [\omega + \epsilon \, \beta,\Phi] \nonumber \\
+ [\omega,[\beta,\Phi]] - [\beta,[\omega,\Phi]] & = & 0 \, . ~~~ \label{cond2}
\end{eqnarray}
If one considers the trivial solution described at the end of Subsection \ref{g0d0}
with $q=\beta$, uses $[\beta,\omega]= \omega + \epsilon \, \beta$ in (\ref{cond2})
and the Jacobi identity, one gets $\omega-[\omega,\beta]=0$ which yields
$\omega = - \epsilon \, \beta /2$. Then using the Jacobi identity in (\ref{cond1})
implies
\[ [\beta,\sigma] + [\dot{\beta},\Phi] + [[\sigma,\beta],\Phi] =0 \]
and using $\dot{\beta} + [\sigma_{0},\beta]=0$ gives $[\beta,\sigma]=[\beta,\sigma_{0}]=0$
or $\sigma_{0}= c \beta(\tau)$ for $c$ an arbitrary real constant, independent of
the choice of the gauge potential $\Phi$. However, this in turn implies that
$\dot{\beta}=0$ or $\beta=\,$constant, the Trautman solution. Hence, once again this
trivial solution is gauge equivalent to the Trautman solution.

Notice that in the general case when $\sigma=\sigma(\tau,a)$ and
$\omega=\omega(\tau,a)$, it is not easy to find a simultaneous solution
to both the conditions (\ref{gd0eq1}), (\ref{gd0eq2}),
(\ref{gd0eq3}), (\ref{gd0eq4}), (\ref{gd0eq5}), (\ref{gd0eq6}),
coming from the $D^{\mu} F_{\mu\nu}=0$ equations, and the gauge conditions
(\ref{cond1}), (\ref{cond2}) above. Hence this class of exact solutions
is not necessarily gauge equivalent to the Trautman solution.

\subsection{\label{ggensol} General Case}

Remember that the solutions in this class were found by solving the
$D^{\mu} F_{\mu\nu}=0$ YM equations to order $R^{-3}$ and by simultaneously
setting the term of order $R^{-1}$ in $\partial^{\mu} F_{\mu\nu}$, i.e. $S_{0}$
(\ref{s0eq}), to zero. Hence these solutions are {\it approximate} in character and
for that reason we now examine the question of whether these are
``approximately" gauge equivalent to the Trautman solution. So we assume
that the gauge potential $\Phi$ locally has a well defined series expansion
in powers of $1/R$ as ($R \neq 0$)
\[ \Phi = \psi + \frac{1}{R} \varphi + \frac{1}{R^2} \zeta + O(R^{-3}) \]
where the coefficients $\psi, \varphi$ and $\zeta$ are, of course, $R$
independent now and we assume them to be only functions of $\tau$ and $a$.
With these in mind, one can write $X$ and $Y$ in (\ref{gau2}) to order $R^{-3}$
as
\begin{eqnarray*}
X & = & \frac{1}{R} (q-\beta-[\beta,\psi]) \\
& & + \frac{1}{R^2} (-\gamma -[\beta,\varphi] -[\gamma,\psi]) + O(R^{-3}) , \\
Y & = & -\sigma - [\sigma,\psi] + \frac{1}{R} (-\omega -[\sigma,\varphi]-[\omega,\psi]) \\
& & + \frac{1}{R^2} (-\delta-[\sigma,\zeta]-[\omega,\varphi]-[\delta,\psi]) + O(R^{-3}) .
\end{eqnarray*}
Remember that at this stage all the coefficients above are only functions of $\tau$
and $a$. So imposing (\ref{omcon}) and carefully collecting the coefficients of
the powers of $1/R$ to order $R^{-3}$, one gets
\begin{eqnarray*}
\dot{z}^{\mu} \partial_{\mu} (q-\beta-[\beta,\psi]) & = & 0 , \\
\omega + [\sigma,\varphi] + [\omega,\psi]
- a (\gamma + [\beta,\varphi] + [\gamma,\psi]) & & \\
+ \dot{z}^{\mu} \partial_{\mu} (\gamma + [\beta,\varphi] + [\gamma,\psi])
- \epsilon \, (q-\beta-[\beta,\psi]) & = & 0 .
\end{eqnarray*}
One now has to solve these two conditions simultaneously with (\ref{geneq1}),
(\ref{geneq2}), (\ref{geneq3}), (\ref{geneq4}) and (\ref{geneq5}) for gauge
equivalence of this class of solutions to the Trautman solution. Obviously,
this is not an easy task if one is to stay in the most general case and
we conjecture that the class of ``approximate" solutions we found are not
``approximately" gauge equivalent to the Trautman solution.

Hence when one considers the solutions presented in Section \ref{sols} in
their full generality, one can assert that they are not gauge equivalent
to the Trautman solution.

\section{\label{concs} Conclusions}

We have found new solutions to the source free YM field equations which
generalize the LW potential of Trautman. Two of the solutions are exact
whereas one of them is approximate and obtained through a $1/R$ series
expansion in the YM field equations. For each solution the total energy
flux $N_{YM}$ and the total color charge $Q$ have been constrained to
be finite. It has also been shown that the solutions are not gauge
equivalent to the Trautman solution in their most general form.

In \cite{gurses}, Trautman's original solution was shown to exist in the
setting of Robinson-Trautman metrics in General Relativity. After the
seminal work of \cite{bamck}, there has also been an ongoing interest
in the particle like solutions of Einstein-YM theory. It would be
interesting to study Einstein-YM theory in the Kerr-Schild geometry
using the general Ansatz for the YM connection (\ref{vacans}) presented
here.

\appendix
\section{\label{app1} The Explicit Form of $D^{\mu} F_{\mu\nu}$}

In this Appendix, we show explicitly how one obtains the YM field equations
starting with the general Ansatz for the YM connection $A_{\mu}$ as
\[ A_{\mu} = H(R,c_i (\tau,a)) \, \dot{z}_{\mu}
+ G(R,c_i (\tau,a)) \, \lambda_{\mu} \, .\]
$R_{,\mu}$ (\ref{rder}), $a_{,\mu}$ (\ref{asubm}) and the derivative of
$a_1$
\begin{equation}
a_{1,\mu} = \frac{1}{R} z^{(3)}_{\mu} - \frac{a_1}{R} \dot{z}_{\mu}
+ \{ a_2 + \frac{1}{R} (\ddot{z}^{\alpha} \ddot{z}_{\alpha})
- a_1 a + \epsilon \, \frac{a_1}{R} \} \lambda_{\mu} \label{a1der}
\end{equation}
are expressions that are needed in the calculation of $D^{\mu} F_{\mu\nu}$.

After lengthy calculations one obtains that
\[ D^{\mu} \, F_{\mu\nu} = X \, z^{(3)}_{\nu} + Y \, \ddot{z}_{\nu}
+ K \, \dot{z}_{\nu} + L \, \lambda_{\nu} \] where
\begin{widetext}
\begin{eqnarray}
X & = & - \frac{1}{R} H_{,c_{i}} c_{i,a} \; , \label{xeq} \\
Y & = & H^{\prime} + \frac{H}{R} - \frac{1}{R} \lbrace R a H^{\prime}_{,c_{i}} +
H_{,c_{i}}^{\;\; ,c_{i}} \, c_{i,\mu} \dot{z}^{\mu} + a H_{,c_{i}} + \epsilon [H, H_{,c_{i}}]
+ [G,H_{,c_{i}}] + G^{\prime}_{,c_{i}} + [H,G_{,c_{i}}] \rbrace c_{i,a} \nonumber \\
& & - \frac{1}{R} H_{,c_{i}} (\dot{c}_{i,a} + c_{i,aa} (a_1 - a^2) -2 a c_{i,a})
\; , \label{yeq} \\
K & = & (R a - \epsilon) H^{\prime\prime} + (3 a -2 \frac{\epsilon}{R}) H^{\prime}
+ \frac{a}{R} H - G^{\prime\prime} - \frac{2}{R} G^{\prime} + H^{\prime}_{,c_{i}} \, c_{i,\mu} \dot{z}^{\mu} + H_{,c_{i}}^{\;\; ,c_{i}} \, c_{i,\mu} c_{i}^{\;\; ,\mu}
+ H_{,c_{i}} c_{i,\mu}^{\;\;\; ,\mu}  \nonumber \\
& & + [G,H^{\prime}] + 2[G^{\prime},H] + \frac{2}{R} [G,H] + (R a - \epsilon) [H,H^{\prime}]
+ [H,H_{,c_{i}} \, c_{i,\mu} \dot{z}^{\mu}] - [H,[H,G]]
\nonumber \\
& & + \frac{1}{R} H_{,c_{i}} \lbrace a_{1} c_{i,a} + a (\dot{c}_{i,a}
+ c_{i,aa} (a_1 - a^2) -2 a c_{i,a}) \rbrace \nonumber \\
& & + \frac{a}{R} \lbrace R a H^{\prime}_{,c_{i}} + H_{,c_{i}}^{\;\; ,c_{i}} \, c_{i,\mu} \dot{z}^{\mu} + a H_{,c_{i}} + \epsilon [H, H_{,c_{i}}]
+ [G,H_{,c_{i}}] + G^{\prime}_{,c_{i}} + [H,G_{,c_{i}}] \rbrace \, c_{i,a} \; , \label{keq} \\
L & = & R a G^{\prime\prime} + 2 G^{\prime}_{,c_{i}} \, c_{i,\mu} \dot{z}^{\mu}
+ G_{,c_{i}}^{\;\; ,c_{i}} \, c_{i,\mu} c_{i}^{\;\; ,\mu} + G_{,c_{i}} c_{i,\mu}^{\;\;\; ,\mu}
- \frac{2}{R} \, G_{,c_{i}} \, c_{i,\mu} \dot{z}^{\mu} - a_{1} (H + R H^{\prime})
\nonumber \\
& & + a (R a - \epsilon) (\frac{H}{R} - R H^{\prime\prime} - H^{\prime})
- (R a - \epsilon) H^{\prime}_{,c_{i}} \, c_{i,\mu} \dot{z}^{\mu}
+ (2 R a - \epsilon) [H^{\prime},G] + [H_{,c_{i}} \, c_{i,\mu} \dot{z}^{\mu},G]
\nonumber \\
& & + (R a + \epsilon) [H,G^{\prime}] + 2 [H,G_{,c_{i}} \, c_{i,\mu} \dot{z}^{\mu}]
+ a [H,G] + \, \epsilon \, [H,[H,G]] + [G,G^{\prime}] + [G,[H,G]]
- \epsilon \, (R a - \epsilon) [H,H^{\prime}] \nonumber \\
& & - \lbrace R a H^{\prime}_{,c_{i}} + H_{,c_{i}}^{\;\; ,c_{i}} \, c_{i,\mu} \dot{z}^{\mu}
+ a H_{,c_{i}} + \epsilon [H, H_{,c_{i}}] + [G,H_{,c_{i}}] + G^{\prime}_{,c_{i}}
+ [H,G_{,c_{i}}] \rbrace (\dot{c}_{i}
+ c_{i,a} (a_{1} - a^2 + \epsilon \, \frac{a}{R})) \nonumber \\
& & - H_{,c_{i}} \lbrace c_{i,a} (a_2 + \frac{1}{R} (\ddot{z}^{\alpha} \ddot{z}_{\alpha})
- a_1 a + \epsilon \, \frac{a_1}{R}) + (a_{1} - a^2 + \epsilon \, \frac{a}{R})
(\dot{c}_{i,a} + c_{i,aa} (a_1 - a^2) -2 a c_{i,a}) \nonumber \\
& & ~~~~~~~~ + \ddot{c}_{i} + \dot{c}_{i,a} (a_1 - a^2) \rbrace \; . \label{leq}
\end{eqnarray}
\end{widetext}

\section{\label{app2} The Explicit Form of $\partial^{\mu} F_{\mu\nu}$}

In this Appendix we give the explicit form of $\partial^{\mu} F_{\mu\nu}$
that is needed in the definition of a total color charge.
Following steps similar to those of Appendix \ref{app1}, one finds that
\[ \partial^{\mu} \, F_{\mu\nu} = B \, z^{(3)}_{\nu} + C \, \ddot{z}_{\nu}
+ P \, \dot{z}_{\nu} + S \, \lambda_{\nu} \] where
\begin{widetext}
\begin{eqnarray}
B & = & - \frac{1}{R} H_{,c_{i}} c_{i,a} \; , \\
C & = & H^{\prime} + \frac{H}{R} - \frac{1}{R} \lbrace R a H^{\prime}_{,c_{i}} +
H_{,c_{i}}^{\;\; ,c_{i}} \, c_{i,\mu} \dot{z}^{\mu} + a H_{,c_{i}}
+ G^{\prime}_{,c_{i}} \rbrace c_{i,a} - \frac{1}{R} H_{,c_{i}}
(\dot{c}_{i,a} + c_{i,aa} (a_1 - a^2) -2 a c_{i,a}) \; , \\
P & = & (R a - \epsilon) H^{\prime\prime} + (3 a -2 \frac{\epsilon}{R}) H^{\prime}
+ \frac{a}{R} H - G^{\prime\prime} - \frac{2}{R} G^{\prime} + H^{\prime}_{,c_{i}} \, c_{i,\mu} \dot{z}^{\mu} \nonumber \\
& & + H_{,c_{i}}^{\;\; ,c_{i}} \, c_{i,\mu} c_{i}^{\;\; ,\mu}
+ H_{,c_{i}} c_{i,\mu}^{\;\;\; ,\mu} + [G,H^{\prime}] + [G^{\prime},H] + \frac{2}{R} [G,H] \nonumber \\
& & + \frac{a}{R} \lbrace R a H^{\prime}_{,c_{i}} +
H_{,c_{i}}^{\;\; ,c_{i}} \, c_{i,\mu} \dot{z}^{\mu} + a H_{,c_{i}}
+ G^{\prime}_{,c_{i}} \rbrace \, c_{i,a} + \frac{1}{R} H_{,c_{i}}
\lbrace a_{1} c_{i,a} + a (\dot{c}_{i,a}
+ c_{i,aa} (a_1 - a^2) -2 a c_{i,a}) \rbrace\; , \\
S & = & R a G^{\prime\prime} + 2 G^{\prime}_{,c_{i}} \, c_{i,\mu} \dot{z}^{\mu}
+ G_{,c_{i}}^{\;\; ,c_{i}} \, c_{i,\mu} c_{i}^{\;\; ,\mu} + G_{,c_{i}} c_{i,\mu}^{\;\;\; ,\mu}
- \frac{2}{R} \, G_{,c_{i}} \, c_{i,\mu} \dot{z}^{\mu} - a_{1} (H + R H^{\prime})
\nonumber \\
& & + a (R a - \epsilon) (\frac{H}{R} - R H^{\prime\prime} - H^{\prime})
- (R a - \epsilon) H^{\prime}_{,c_{i}} \, c_{i,\mu} \dot{z}^{\mu}
+ R a [H^{\prime},G] + R a [H,G^{\prime}] + [H_{,c_{i}} \, c_{i,\mu} \dot{z}^{\mu},G]
\nonumber \\
& & + [H,G_{,c_{i}} \, c_{i,\mu} \dot{z}^{\mu}]
- \lbrace R a H^{\prime}_{,c_{i}} + H_{,c_{i}}^{\;\; ,c_{i}} \, c_{i,\mu} \dot{z}^{\mu}
+ a H_{,c_{i}} + G^{\prime}_{,c_{i}} \rbrace (\dot{c}_{i}
+ c_{i,a} (a_{1} - a^2 + \epsilon \, \frac{a}{R})) \nonumber \\
& & - H_{,c_{i}} \lbrace c_{i,a} (a_2 + \frac{1}{R} (\ddot{z}^{\alpha} \ddot{z}_{\alpha})
- a_1 a + \epsilon \, \frac{a_1}{R}) + (a_{1} - a^2 + \epsilon \, \frac{a}{R})
(\dot{c}_{i,a} + c_{i,aa} (a_1 - a^2) -2 a c_{i,a}) \nonumber \\
& & ~~~~~~~~ + \ddot{c}_{i} + \dot{c}_{i,a} (a_1 - a^2) \rbrace \; .
\end{eqnarray}
\end{widetext}

\begin{acknowledgments}

I would like to thank Prof. Metin G{\" u}rses for helpful discussions and his
careful reading of this manuscript.

\end{acknowledgments}

\end{document}